%% file: sample-sigconf-authordraft.tex
\documentclass[sigconf,nonacm , dvipsnames]{acmart}

\usepackage{booktabs}
\usepackage{multirow}
\usepackage{pifont}
\usepackage{tcolorbox}
\usepackage{listings}
\usepackage[table]{xcolor}
\usepackage{booktabs}
\usepackage{colortbl}

\usepackage{mdframed}
\usepackage{xcolor}

\definecolor{takeawayBg}{RGB}{240,240,240}
\definecolor{takeawayRule}{RGB}{100,100,100}

\mdfdefinestyle{takeawaystyle}{
  backgroundcolor=gray!12,
  linecolor=gray!50,
  linewidth=0.8pt,
  roundcorner=5pt,
  innerleftmargin=8pt,
  innerrightmargin=8pt,
  innertopmargin=7pt,
  innerbottommargin=7pt,
  skipabove=6pt,
  skipbelow=6pt,
}

\AtBeginDocument{%
  }

\begin{document}

\title{When Prompt Under-Specification Improves Code Correctness: An Exploratory Study of Prompt Wording and Structure Effects on LLM-Based Code Generation}


\author{Amal Akli}
\orcid{}
\affiliation{%
  \institution{University of Luxembourg}
  \city{}
  \country{Luxembourg}}
\email{amal.akli@uni.lu} 

\author{Mike Papadakis}
\affiliation{%
  \institution{University of Luxembourg}
  \city{}
  \country{Luxembourg}}
\email{michail.papadakis@uni.lu}

\author{Maxime Cordy}
\orcid{}
\affiliation{%
  \institution{University of Luxembourg}
  \city{}
  \country{Luxembourg}}
\email{maxime.cordy@uni.lu}

\author{Yves Le Traon}
\affiliation{%
  \institution{University of Luxembourg}
  \city{}
  \country{Luxembourg}}
\email{Yves.LeTraon@uni.lu}

\renewcommand{\shortauthors}{Akli et al.}

\begin{abstract}
 Large language models (LLMs) are increasingly used for code generation, yet the correctness of their outputs depends not only on model capability but also on how tasks are specified. Prior studies demonstrate that small changes in natural language prompts, particularly under-specification can substantially reduce code correctness; however, these findings are largely based on minimal-specification benchmarks such as HumanEval and MBPP, where limited structural redundancy may exaggerate sensitivity. In this exploratory study, we investigate how prompt structure, task complexity, and specification richness interact with LLM robustness to prompt mutations. We evaluate 10 models, ranging from small (6–7B) to large open-source (15–34B) and cutting-edge reasoning models, across HumanEval and the structurally richer LiveCodeBench. Our results reveal that robustness is not a fixed property of LLMs but is highly dependent on prompt structure: the same under-specification mutations that degrade performance on HumanEval have near-zero net effect on LiveCodeBench due to redundancy across descriptions, constraints, examples, and I/O conventions. Surprisingly, we also find that prompt mutations can improve correctness. In LiveCodeBench, under-specification often breaks misleading lexical or structural cues that trigger incorrect retrieval-based solution strategies, leading to correctness improvements that counterbalance degradations. Manual analysis identifies consistent mechanisms behind these improvements, including the disruption of over-fitted terminology, removal of misleading constraints, and elimination of spurious identifier triggers. Overall, our study shows that structurally rich task descriptions can substantially mitigate the negative effects of under-specification and, in some cases, even enhance correctness. We outline categories of prompt modifications that positively influence the behavior of LLM code-generation, offering practical insights into the construction of robust coding prompts.
\end{abstract}

\maketitle

\input{introduction}

\input{Dataset}
\input{Experimental_Setup}
\input{results}

\input{Related_work}

\input{threats}
\input{conclusion}




\clearpage
\bibliographystyle{ACM-Reference-Format}
\bibliography{sample-base}

\appendix

\end{document}

%% file: introduction.tex
\section{Introduction}

\begin{figure*}[t]
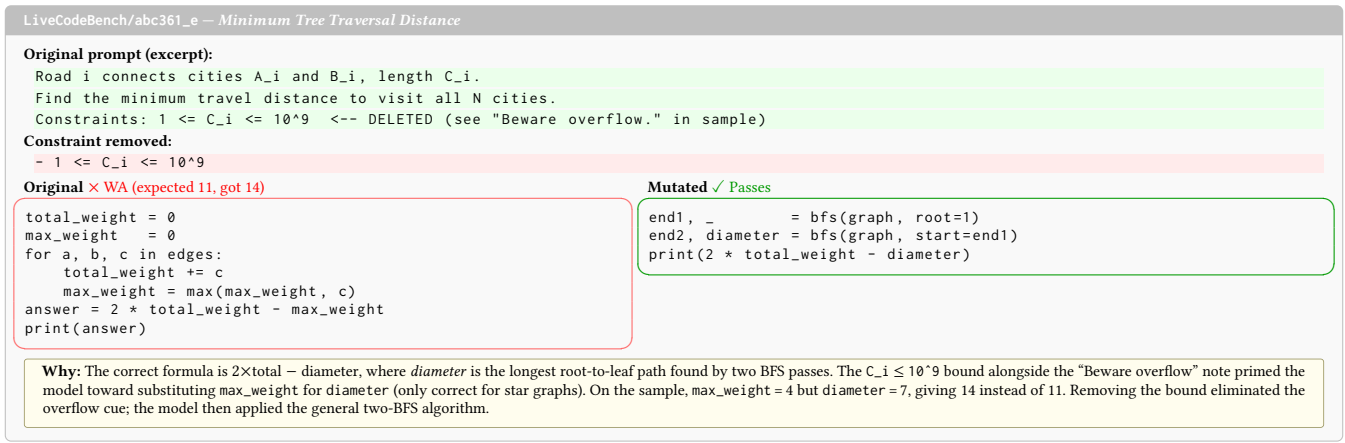

\vspace{-0.5em}
\centering
\scriptsize
\begin{tcolorbox}[
    colback=gray!5, colframe=gray!50,
    title={\texttt{LiveCodeBench/abc361\_e} --- \textit{Minimum Tree Traversal Distance}},
    fonttitle=\scriptsize\bfseries,   
    boxrule=0.4pt, arc=2pt,
    left=4pt, right=4pt, top=2pt, bottom=2pt  
]
\textbf{Original prompt (excerpt):}
\begin{lstlisting}[basicstyle=\scriptsize\ttfamily, backgroundcolor=\color{green!8},
                   frame=none, xleftmargin=4pt, aboveskip=1pt, belowskip=1pt]
Road i connects cities A_i and B_i, length C_i.
Find the minimum travel distance to visit all N cities.
Constraints: 1 <= C_i <= 10^9  <-- DELETED (see "Beware overflow." in sample)
\end{lstlisting}
\textbf{Constraint removed:}
\begin{lstlisting}[basicstyle=\scriptsize\ttfamily, backgroundcolor=\color{red!8},
                   frame=none, xleftmargin=4pt, aboveskip=1pt, belowskip=2pt]
- 1 <= C_i <= 10^9
\end{lstlisting}
\begin{minipage}[t]{0.46\linewidth}
\textbf{Original} \textcolor{red}{$\times$ WA (expected 11, got 14)}
\begin{lstlisting}[basicstyle=\scriptsize\ttfamily, frame=single,
                   frameround=tttt, rulecolor=\color{red!60},
                   aboveskip=1pt, belowskip=0pt]
total_weight = 0
max_weight   = 0
for a, b, c in edges:
    total_weight += c
    max_weight = max(max_weight, c)
answer = 2 * total_weight - max_weight
print(answer)
\end{lstlisting}
\end{minipage}
\hfill
\begin{minipage}[t]{0.52\linewidth}
\textbf{Mutated} \textcolor{green!60!black}{$\checkmark$ Passes}
\begin{lstlisting}[basicstyle=\scriptsize\ttfamily, frame=single,
                   frameround=tttt, rulecolor=\color{green!60!black},
                   aboveskip=1pt, belowskip=0pt]
end1, _        = bfs(graph, root=1)
end2, diameter = bfs(graph, start=end1)
print(2 * total_weight - diameter)
\end{lstlisting}
\end{minipage}
\begin{tcolorbox}[colback=yellow!8, colframe=yellow!50!black,
                  boxrule=0.3pt, arc=1pt, left=4pt, right=4pt,
                  top=1pt, bottom=1pt, before upper={\vspace{-2pt}}]  
\scriptsize                                          
\textbf{Why:} The correct formula is $2{\times}\text{total} - \text{diameter}$, where
\emph{diameter} is the longest root-to-leaf path found by two BFS passes.
The \texttt{C\_i\,$\leq$\,10\^{}9} bound alongside the ``Beware overflow'' note
primed the model toward substituting \texttt{max\_weight} for \texttt{diameter}
(only correct for star graphs). On the sample, \texttt{max\_weight}\,=\,4 but
\texttt{diameter}\,=\,7, giving 14 instead of 11. Removing the bound eliminated
the overflow cue; the model then applied the general two-BFS algorithm.
\end{tcolorbox}
\end{tcolorbox}
\vspace{-2.0em}
\caption{A LiveCodeBench example where removing a constraint \emph{improves} generated code:
the constraint primed the model toward an incorrect algorithm; its removal forced
reasoning from problem structure.}
\label{fig:us_motivating_example}
\vspace{-0.5em}
\end{figure*}


Large language models (LLMs) such as CodeLlama~\cite{roziere2023codellama}, DeepSeekCoder~\cite{guo2024deepseek}, Qwen2.5Coder~\cite{hui2024qwen25coder}, and proprietary API-based models~\cite{openai2025gpt5,anthropic2024claude} are now widely used for code generation in real-world development workflows~\cite{generation2026survey}. Despite these advances, the correctness of generated code remains highly variable, and understanding the factors that influence this variability is critical for both practitioners and researchers.

A key factor is the \emph{natural-language specification} used to describe the task. In contrast to traditional program synthesis, where specifications are formal and precise, LLM-based code generation relies on prompts that are often incomplete, ambiguous, or inconsistently structured. Prior work shows that even minor changes in wording can significantly affect correctness~\cite{formatspread, zhu2023promptbench, NLPerturbator, mastropaolo2023robustness, PereraAKC23}, and that under-specification (i.e., missing or weakened requirements) can substantially degrade performance~\cite{larbi2025prompts}. These findings have led to a prevailing assumption that more precise and more complete prompts are necessary for reliable code generation.

However, existing evidence is almost exclusively derived from minimal-specification benchmarks such as HumanEval and MBPP. These benchmarks express tasks through short, single-docstring descriptions with limited structural redundancy, making them highly sensitive to perturbations. They also exhibit contamination issues, which may result in fallacious experimental results~\cite{yang2023rethinking}. In contrast, real-world programming tasks typically include multiple complementary specification layers \cite{WhiteFGBSSMK23}, including descriptions, constraints, input/output formats, and examples. These components often encode overlapping information, suggesting that robustness to prompt imperfections may depend not only on wording, but also on the \emph{structure and richness of the specification}.

This observation motivates us to study a fundamental question: \emph{how prompt wording and structure relate to code correctness in LLM-based code generation?} Addressing this question requires moving beyond minimal benchmarks and considering both the structure of prompts and the nature of the words and specificity they contain.

In this paper, we conduct an exploratory study of how prompt wording, structure, and task complexity interact with code correctness in LLM-based code generation. We systematically introduce prompt mutations by altering original task descriptions, in two benchmark of contrasting complexities: HumanEval (HE -- simple task descriptions) and LiveCodeBench (LCB -- detailed, multi-layered descriptions). We evaluate 10 models spanning small and large open-source models as well as state-of-the-art proprietary models. We measure the effect of under specification based on aggregate performance (Pass@1) and finer analysis of test result transitions (Pass$\rightarrow$Fail and Fail$\rightarrow$Pass). We also performed a manual root-cause analysis of cases where prompt mutations surprisingly lead to improvements. Our results reveal a more nuanced picture of LLM robustness than previously reported, via three main findings.

\textbf{Finding 1. Robustness to under-specification is governed by prompt structure rather than by model size or architecture.}
Mutations that cause substantial performance drops on HE have near-zero net impact on LCB, where multi-layered specifications provide redundancy across descriptions, constraints, examples, and I/O formats. Attention patterns further shows that models rely heavily on these auxiliary components, which act as fallback signals when parts of the prompt are degraded.

\textbf{Finding 2. Under-specification both harms and improves correctness, with these effects canceling out in rich prompts.}
On LCB, improvements and degradations occur in nearly equal proportion (Fail$\rightarrow$Pass/Pass$\rightarrow$Fail = 0.99), whereas degradations dominate on HE (0.40). We identify 69 LCB tasks that consistently improve pass rate under prompt mutations across multiple models, indicating that certain elements of the original specification can hinder correct code generation.

\textbf{Finding 3. Improvements arise when under-specification disrupts misleading retrieval cues.}
Specific words, identifiers, or constraint formulations can anchor models to incorrect, memorized solution patterns. Removing or weakening these cues reduces overfitting to surface forms and encourages reasoning from the underlying problem structure, often leading to simpler and more accurate solutions.

Overall, our findings challenge the assumption that more detailed specifications necessarily yield better code generation. Instead, they highlight the importance of prompt structure and the complex role of natural-language cues in shaping model behavior. By identifying both harmful and beneficial prompt characteristics, this work provides actionable insights for constructing robust prompts and calls for a re-evaluation of how code generation systems are benchmarked and used in practice.

\section{Exploratory study objectives}
We seek to understand how prompt structure, complexity and wording characteristics relate to code correctness. We make a manual exploratory study that looks into the prompt pairs of under-specified and well-specified task descriptions and related them with the generated code correctness. In particular, we are trying to understand what is causing the LLM to generate correct and incorrect code.  

Figure \ref{fig:us_motivating_example} shows an example pair of a well-formed and under-specified prompt that leads to incorrect and correct LLM-generated code. It is surprising to see that ``less is more,'' the removal of an important constraint leads to correct code. Through our exploratory study we hypothesize that this is caused by the specific wording used that wrongly associates the sought solution with a wrong concept, thereby preventing the model from reasoning about the whole information provided. We exploratory come to our explanations and observations by identifying the same patterns across multiple different prompt pairs.  As such our study is exploratory by nature and aims to increase our understanding of what could improve generated code correctness, leading to the formulation of concrete observations that could help the research community make a better use of existing tools and could lead stronger code generation tools. 

In particular, we summarize our objectives-questions:

 \textbf{What is the relation between code correctness and task complexity and specification richness in the presence and absence of prompt mutations?}     
    We evaluate 10 code generation LLMs, 
    including 2 reasoning models (GPT-5-mini and Claude-Sonnet-4),  on HumanEval (easy tasks ) and LiveCodeBench (complex tasks). We consider 3 mutant types (LV, US, SF) and measure Pass@1 changes to determine whether the effect of the mutations relate to the complexity and richness of the underlying tasks.

 \textbf{ Does prompt under-specification uniformly degrade code generation performance ?} 
    The near-zero net effect observed on LiveCodeBench in RQ1 could reflect either insensitivity to mutations or the cancellation of two opposing forces. 
    To distinguish between these explanations, we analyze Pass@1 changes and transitions, solutions that passed before mutation and fail after (P$\to$F), and solutions that failed before and pass after (F$\to$P), and check if the observed balance is systematic. 

 \textbf{What are the key characteristics of prompts that lead to correct code as opposed to those that lead to incorrect code for the same tasks?}     
    Having established that mutations cause a consistent correctness improvement on LCB, we investigate what is causing it. We manually analyze F$\to$P transitions on LiveCodeBench and construct a root-cause taxonomy, identifying how specific elements of the original specification (vocabulary, identifier names, and constraint lines) steer models toward incorrect solutions and how their removal or softening steers them toward correct ones.

%% file: Dataset.tex
\section{Prompts used}

To study how under-specification affect code correctness, we systematically mutate task descriptions using three types of transformations. The resulting 3,651 mutated instances are validated by an independent LLM judge, with a stratified random sample being reviewed by the authors of the paper.

\subsection{Selected Benchmarks}

We select two benchmarks that are structurally contrastive along the dimensions most relevant to prompt sensitivity: specification richness and task complexity.

\textbf{HumanEval}~\cite{humaneval}  consists of 164 hand-written Python programming problems. Each problem communicates its specification exactly through a function signature with a short docstring (~231 characters average), validated by unit tests. There are no separate constraints sections, no structured sample I/O, no input format specifications. Previous studies report that HumanEval suffers from benchmark contamination, with 8 to 18\% direct overlap with training datasets \cite{yang2023rethinking}, suggesting that many HumanEval problems can be memorized from publicly available repositories.

\textbf{LiveCodeBench}~(LCB)~\cite{livecodebench2024} collects problems from three competitive programming platforms (LeetCode, AtCoder, Codeforces), each time stamped at publication. follow a full competitive programming format, each problem contains (1) full prose description, (2) explicit constraints section with value ranges, (3) multiple sample I/O pairs with explanations, (4) input/output format specification. These channels encode overlapping information. This benchmark is contamination-free by design. Problems are time-stamped from weekly contests on LeetCode, Codeforces, and AtCoder. We sample latest 1{,}055 problems that range from easy to hard difficulty.

 
\subsection{Mutation of original benchmarks }

We create under-specified prompts by mutating original prompts. Thus, each description yields three mutated variants, one per mutation type.  In cases where the original descriptions are too short or have no constraint to delete, no mutation is produced. All mutations are applied exclusively at the prompt level; reference solutions, test cases, and evaluation harnesses remain untouched — ensuring that any Pass@1 change can be attributed solely to the modified description.

\subsubsection{Mutation Type Selection}

 Our three mutation types are grounded in prior empirical findings. Token-level perturbations and lexical ambiguity have been shown to introduce substantial variance in code generation correctness~\cite{recode,NLPerturbator,larbi2025prompts}. Incomplete or missing requirements, even when the description remains grammatically well-formed, steer models toward solutions that satisfy the syntax but miss the intended semantics~\cite{mu2024clarifygpt,SpecFix,TiCoder,HamidiKP25}. Formatting noise, despite being purely surface-level, can produce measurable correctness swings in open-source models~\cite{formatspread,recode}. Each class targets a distinct channel through which a natural-language specification can fail, making the three types complementary rather than redundant. Table~\ref{tab:mutation_examples} provides representative examples of each mutation type applied to problems from the two benchmarks.

\subsubsection{Mutation Rules}
\paragraph{Lexical Vagueness (LV)}
LV mutations reduce the specificity of a task by replacing precise, task-relevant terms with semantically broader or less informative alternatives, while preserving all explicit constraints. This type of transformation is closely related to paraphrasing and synonym-substitution perturbations studied by Wang et al.~\cite{recode} and Chen et al.~\cite{NLPerturbator}. In practice, an LV mutation can: replace precise terms or verbs with broader synonyms (e.g. ``sort'' becomes ``arrange''); rename function and parameter identifiers into less informative names (e.g., ``delimiter'' becomes ``filler''); generalize or remove type annotations.

\paragraph{Under-Specification (US)}
US mutations remove a single explicit constraint from the prompt, resulting in a description that remains grammatically well-formed but becomes semantically under-specified. This design is motivated by empirical evidence showing that incomplete requirements are a primary source of intent mismatch in LLM-based code generation~\cite{larbi2025prompts}. To ensure consistent under-specification, we remove only: bounds/threshold values; ordering rules; input preconditions; output edge-cases description; formal constraints; error and exception handling requirements; output type and format constraints.

\paragraph{Syntax and Formatting (SF)}
SF mutations introduce surface-level noise into the prompt without altering its semantic content, targeting the susceptibility of LLMs to low-level textual  corruption~\cite{formatspread}. Sclar et al.~\cite{formatspread} demonstrated that 
format variations alone can have a significant impact, motivating this mutation type. Each SF mutation can create noise at different levels: tokens; delimiters, brackets and colons; indentation and whitespaces; example-block formatting.

\begin{table*}[ht]
\centering
\small
\vspace{-0.4em}
\caption{Examples of mutation strategies applied to \textsc{HumanEval} and \textsc{LiveCodeBench}. \textbf{US} (Under-Specification) removes a constraint from the docstring; \textbf{LV} (Lexical Vagueness) paraphrases using different vocabulary and renames identifiers; \textbf{SF} (Syntax/Formatting) introduces typographical errors and syntax noise into both the signature and docstring.}
\vspace{-0.7em}
\label{tab:mutation_examples}
\renewcommand{\arraystretch}{1.4}
\begin{tabular}{p{1.1cm} p{2.0cm} p{10.5cm}}
\toprule
\textbf{Element} & \textbf{Mutation type} & \textbf{Original prompt and its mutations} \\
\midrule

\multirow{4}{*}{LCB/2811}
& Original & \textit{``You are given two integers, $n$ and $k$.''} \newline
           \textit{``An array of \underline{distinct positive}  integers is called a k-avoiding array if there does not exist any pair of distinct elements that sum to $k$. Return the minimum possible sum of a k-avoiding array of length $n$.''} \newline
           \texttt{Input: n=5, k=4} $\Rightarrow$ \texttt{Output: 18} \quad \texttt{Input: n=2, k=6} $\Rightarrow$ \texttt{Output: 3} \\
& US       & \textit{``You are given two integers, $n$ and $k$.''} \newline
           \textit{``An array of integers is called a k-avoiding array if there does not exist any pair of distinct elements that sum to $k$. Return the minimum possible sum of a k-avoiding array of length $n$.''} \newline
           \texttt{Input: n=5, k=4} $\Rightarrow$ \texttt{Output: 18} \quad \texttt{Input: n=2, k=6} $\Rightarrow$ \texttt{Output: 3} \\
& LV       & \textit{``A \underline{list} of \underline{different positive numbers} is considered a k-avoiding list if there aren't any two different members that \underline{add up to} $k$. Return the \underline{smallest possible total} of a k-avoiding list \underline{containing $n$ entries}.''} \newline
           \texttt{Input: n=5, k=4} $\Rightarrow$ \texttt{Output: 18} \quad \texttt{Input: n=2, k=6} $\Rightarrow$ \texttt{Output: 3} \\
& SF       & \texttt{\underline{INPUT;\{"specific\_prompt":"}You are given two \underline{integres}, $n$ and $k$.} \newline
           \textit{``An array of distinct positive integers is called a k-avoiding array if there does not exist any pair of distinct elements that sum to. Return the minimum possible sum of a Return the minimum possible sum \ldots''} \newline
           \texttt{Input: n=5, k=4} $\Rightarrow$ \texttt{Output: 18} \quad \texttt{\underline{Constraints} [no colon] \ldots \underline{**INVALID****OCUMENT**\}**}} \\

\midrule
           
\multirow{4}{*}{HE/5}
& Original & \texttt{def intersperse(numbers: List[int], delimeter: int) -> List[int]:} \newline
           \textit{``Insert a number `delimeter' \underline{between every two consecutive elements} of input list `numbers'.''} \newline
           \texttt{>>> intersperse([1, 2, 3], 4)} \; $\Rightarrow$ \; \texttt{[1, 4, 2, 4, 3]} \\
& US       & \texttt{def intersperse(numbers: List[int], delimeter: int) -> List[int]:} \newline
           \textit{``Insert a number `delimeter' \underline{into} the input list `numbers'.''} \newline
           \texttt{>>> intersperse([1, 2, 3], 4)} \; $\Rightarrow$ \; \texttt{[1, 4, 2, 4, 3]} \\
& LV       & \texttt{def \underline{place\_between}(\underline{items}, \underline{filler}):} \newline
           \textit{``Put a value `filler' between \underline{neighboring entries} of the provided \underline{sequence}.''} \newline
           \texttt{>>> place\_between([1, 2, 3], 4)} \; $\Rightarrow$ \; \texttt{[1, 4, 2, 4, 3]} \\
& SF       & \texttt{def intersperse(numbers: List[int\underline{)}, \underline{delimter}: int) -> List[int\underline{)}} \newline
           \texttt{\underline{'''} Insert a number `\underline{delimter}' between every two consecutive elements of input list `numbers' \ldots} \\

\bottomrule
\end{tabular}
\end{table*}

\subsubsection{Mutants Generation}

Mutations are generated using \texttt{gpt-5-mini} via the OpenAI Batch API~\cite{openai2025gpt5}. A structured prompt is used for each mutation type, specifying transformation rules and constraining the model output. The generation process is iterated, with rule refinements applied as needed, until the automated judge (described below) assigns compliance scores above 85\% for all types of mutations and benchmarks\footnote{All generation prompts are available in the replication package. \url{https://github.com/Amal-AK/PromptAnalysis}}.

\subsection{Validation of Mutated Benchmarks}

We validate mutations using an independent LLM judge followed by manual annotation of a stratified sample.

\subsubsection{Automated Quality Assessment}
\textbf{Judge model}.
We use \textit{Qwen2.5-Coder-32B-Instruct}~\cite{hui2024qwen25codertechnicalreport} as the judge. It is architecturally distinct from the mutation generator (\texttt{gpt-5-mini}), limiting collusion risk~\cite{llmjudge}; it runs locally, ensuring full reproducibility; and its instruction-following capability is sufficient for our binary-judgment protocol. The model runs in float16 with greedy decoding, outputting a single JSON object (\texttt{{"score": 0}} or \texttt{{"score": 1}}).

\textbf{Evaluation criteria.}
Each mutant is assessed for compliance and naturalness, both framed as binary yes/no questions. Compliance is type-specific: for LV, the judge asks whether only vocabulary or identifiers were made less precise without altering the task; for SF, whether only typographical or formatting noise was introduced; for US, whether exactly one requirement was removed and nothing else changed. Naturalness asks whether the resulting description reads as something a real developer might plausibly write. The scores reported are the proportion of "yes" answers across all mutants of a given benchmark-mutation pair.

\textbf{Evaluation results.} LV and SF achieve near-perfect compliance (0.99--1.00) and high naturalness (0.96--0.99) across both benchmarks. US compliance is slightly lower (0.89 on HumanEval, 0.85 on LiveCodeBench), as structural constraints are stricter in LCB and removing them can produce unusually terse problem statements, reflected in a naturalness drop to 0.61 on LCB for US mutations.

\subsubsection{Manual Inspection}

To ground the automated scores in human judgment, three researchers independently annotated a stratified random sample of 100 mutants covering all mutation types and benchmarks, approximately 11 instances per benchmark-mutation pair, using the same binary compliance and naturalness criteria as the judge. Disagreements were resolved by majority vote. Annotators agreed with the judge on 97\% of compliance assessments and 86\% of naturalness assessments, confirming that the automated scores are a reliable proxy for human evaluation and that the reported quality figures can be trusted

%% file: Experimental_Setup.tex
\section{Experimental Setup}

We describe below our general experimental setup used throughout our study. All scripts and data are publicly available. \footnote{\url{https://anonymous.4open.science/r/PromptAnalysis/}}

\subsection{ Models Under Study}

We evaluate a diverse set of code-generation models to ensure our findings are not artifacts of a single architecture or training regime. The models are grouped into three levels.

\textbf{Small open-source models} ($\sim$7B parameters): Qwen2.5-Coder-7B~\cite{hui2024qwen25coder}, DeepSeek-Coder-6.7B~\cite{guo2024deepseek}, and CodeLlama-7B~\cite{roziere2023codellama}. These models are representative of deployable, resource-constrained settings and are widely used as baselines in the literature.

\textbf{Large open-source models} (15B--34B parameters): Qwen2.5-Coder-32B~\cite{hui2024qwen25coder}, DeepSeek-Coder-33B~\cite{guo2024deepseek}, CodeLlama-34B~\cite{roziere2023codellama}, Codestral-22B~\cite{mistral2024codestral}, and StarCoder2-15B~\cite{lozhkov2024starcoder2}. This tier allows us to examine whether larger capacity yields greater robustness to prompt mutations, independent of proprietary training data or alignment procedures.

\textbf{Closed-source API models}:  GPT-5 mini~\cite{openai2025gpt5}, and Claude Sonnet 4~\cite{anthropic2024claude}. These represent the current state-of-the-art in code generation and serve as upper-bound references. Their inclusion enables a direct comparison between reasoning models \cite{WeiWSL22} and open-source alternatives under identical evaluation conditions.

Collectively, this selection spans five model families, two provider categories (open-source and proprietary), and two scale regimes, enabling controlled cross-family and cross-scale comparisons. 

We use greedy decoding (temperature $= 0$) to ensure deterministic model outputs. Each model produces exactly one outcome per problem, and re-running the same experiment yields identical results. API models are queried via the OpenAI and Anthropic batch APIs, while open-source models are served locally on NVIDIA A100 GPUs in float16 precision. Generated solutions are extracted via markdown fenced code block parsing, with fallback to model-specific delimiters (e.g., CodeLlama's \texttt{[PYTHON]} tags). Each solution is executed in an isolated process with a 20-second timeout to safely handle infinite loops and runtime errors in generated code.

\subsection{Metrics}

Pass@1~\cite{chen2021codex} is used as the primary correctness metric, as it reflects the real-world scenario where a developer accepts the model's first suggestion. This metric is the de facto standard in code generation evaluation~\cite{chen2021codex, mbpp} and enables direct comparison with prior work.

Additionally, we complement Pass@1 with a transition-based analysis that examines test outcomes (pass or fail) before and after mutation. Thus, we track (i) \emph{Pass$\rightarrow$Fail} transitions, where a previously correct solution becomes incorrect after mutating the prompt, and (ii) \emph{Fail$\rightarrow$Pass} transitions, where a previously incorrect solution becomes correct. This finer-grained analysis allows us to distinguish between uniform degradation and more nuanced effects where improvements and regressions coexist and potentially cancel out at the aggregate level.

%% file: results.tex
\section{Experimental Results}

\subsection{Impact of mutations}

Table~\ref{tab:all_results} reveals a consistent asymmetry among the results of HumanEval and LiveCodeBench; HumanEval is systematically degraded by all three types of mutations used, while LiveCodeBench remains near-neutral, with the gap continuously widening from SF to LV to US across all model sizes. 

\textbf{The three mutant types induce markedly different levels of degradation on HumanEval, yet remains largely ineffective on LiveCodeBench.} US causes the largest drops on HumanEval (avg. $-11.8\%$), consistent with prior findings that missing constraints are the most consequential source of intent mismatch~\cite{mu2024clarifygpt,SpecFix}, while the same mutant type produces only $-0.9\%$ on LCB. LV produces moderate degradation on HumanEval (avg. $-7.1\%$), compared to a near-zero average of $-0.2\%$ on LCB, and several models record marginal improvements. Table~\ref{tab:lv_ablation} decomposes this effect into docstring paraphrasing ($\Delta_{\text{vocab}}$) and identifier renaming ($\Delta_{\text{name}}$): both contribute independently, but $\Delta_{\text{name}}$  dominates in smaller models (DeepSeek-6.7B: $-9.1\%$ vs. $-3.1\%$ for vocabulary alone), suggesting that these models are disproportionately anchored to function signatures as retrieval cues, a pattern consistent with memorisation-driven processing \cite{CarliniTWJHLRBS21} rather than specification understanding. SF represents a practical lower bound on both benchmarks (avg. $-1.5\%$ on HumanEval, $-0.4\%$ on LCB)~\cite{sclar2024}, with its effect on the LCB sign inconsistent across models.

Execution rates in Table \ref{tab:all_results} exhibit smaller declines across models and mutation types compared to the larger drops observed in Pass@1. This suggests that our mutations mainly affect functional correctness rather than causing syntactic or runtime failures.

\textbf{Model scale provides no consistent protection}. As Table \ref{tab:all_results} shows, the degradation is identical across small, large, and API model groups on HumanEval, with API level reasoning models equally affected (GPT-5-mini and Claude Sonnet~4 both losing $\sim$9.7 \% under US).

\begin{figure}[t]
    \centering
    \vspace{-0.5em}
    \includegraphics[width=1.1\linewidth]{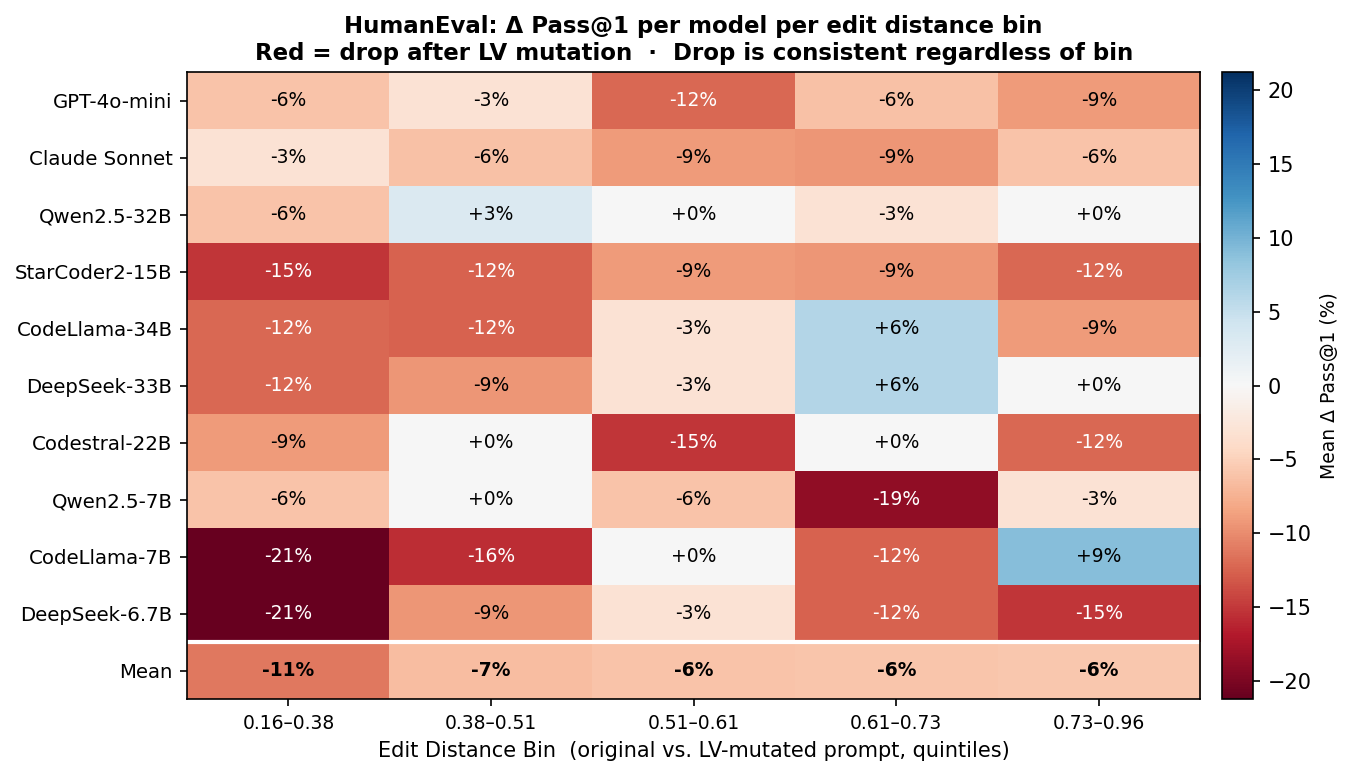}
    \vspace{-0.8em}
     \caption{Pass@1 delta under LV mutations on HumanEval as a function of edit distance between original and mutated prompts (quintile bins). Each cell reports the mean Pass@1 change per model; red indicates degradation.}
     \vspace{-0.5em}
    \label{fig:edit_distance}
\end{figure}

\textbf{The divergence between benchmarks is not explained by mutation magnitude but by specification structure}. Figure~\ref{fig:edit_distance} shows that in HumanEval, drops below LV are uniformly negative across all edit distance bins with no correlation between change magnitude and severity, pointing to memorized surface form dependence rather than noise sensitivity. Figure~\ref{fig:attention} provides the structural explanation: on HumanEval, models concentrate 60--86\% of attention on the description, the only available specification, whereas on LiveCodeBench, 30--50\% of attention falls on I/O format and sample I/O regions that remain intact on mutated prompts.

\begin{mdframed}[style=takeawaystyle]
\textbf{Takeaway.}
Sensitivity to prompt wording is an artifact of specification structure: prompts encoding information across multiple independent regions tend to be robust, suggesting that robustness results reported by previous studies overestimate the fragility of code LLMs when prompts combine descriptions, constraints, and examples. 
\end{mdframed}

\begin{table*}[t]
\centering
\small
\caption{
  Pass@1 and execution rate results across benchmarks and mutation types.
  \textbf{P@1}: pass@1 (\%). \textbf{Ex}: execution rate (\%).
  For each mutation (US, LV, SF), Pass@1 deltas relative to the original are shown in
  {\color{red}red} ($\downarrow$\,drop) and {\color{green!60!black}green} ($\uparrow$\,gain).
}
\label{tab:all_results}
\resizebox{\textwidth}{!}{%
\begin{tabular}{l | rr rr rr rr | rr rr rr rr}
\toprule

& \multicolumn{8}{c|}{\textbf{HumanEval}}
& \multicolumn{8}{c}{\textbf{LiveCodeBench}} \\

\cmidrule(lr){2-9}\cmidrule(lr){10-17}

& \multicolumn{2}{c}{Orig}
& \multicolumn{2}{c}{US}
& \multicolumn{2}{c}{LV}
& \multicolumn{2}{c|}{SF}
& \multicolumn{2}{c}{Orig}
& \multicolumn{2}{c}{US}
& \multicolumn{2}{c}{LV}
& \multicolumn{2}{c}{SF} \\

\cmidrule(lr){2-3}\cmidrule(lr){4-5}\cmidrule(lr){6-7}\cmidrule(lr){8-9}
\cmidrule(lr){10-11}\cmidrule(lr){12-13}\cmidrule(lr){14-15}\cmidrule(lr){16-17}

\textbf{Model}
  & P@1 & Ex & P@1 & Ex & P@1 & Ex & P@1 & Ex
  & P@1 & Ex & P@1 & Ex & P@1 & Ex & P@1 & Ex \\

\midrule
\multicolumn{17}{l}{\textit{Small models ($\leq$7B)}} \\[1pt]

CodeLlama-7B
  & 37.2 & 84.8
  & 29.3\,{\color{red}$\downarrow$7.9} & 90.9
  & 29.3\,{\color{red}$\downarrow$7.9} & 81.7
  & 37.2\,{\color{gray}=}              & 87.8
  &  8.1 &  8.2
  &  8.4\,{\color{green!60!black}$\uparrow$0.3} &  8.5
  &  8.2\,{\color{green!60!black}$\uparrow$0.1} &  8.3
  &  7.8\,{\color{red}$\downarrow$0.3} &  7.9 \\

DeepSeek-Coder-6.7B
  & 72.6 & 97.0
  & 57.3\,{\color{red}$\downarrow$15.3} & 93.9
  & 60.4\,{\color{red}$\downarrow$12.2} & 95.7
  & 68.9\,{\color{red}$\downarrow$3.7}  & 93.9
  & 16.3 & 16.4
  & 16.0\,{\color{red}$\downarrow$0.3} & 16.1
  & 14.8\,{\color{red}$\downarrow$1.5} & 14.9
  & 15.2\,{\color{red}$\downarrow$1.1} & 15.3 \\

Qwen2.5-Coder-7B
  & 82.3 & 97.6
  & 67.7\,{\color{red}$\downarrow$14.6} & 97.0
  & 75.6\,{\color{red}$\downarrow$6.7}  & 95.1
  & 81.7\,{\color{red}$\downarrow$0.6}  & 98.2
  & 20.5 & 20.6
  & 20.3\,{\color{red}$\downarrow$0.2} & 20.4
  & 22.2\,{\color{green!60!black}$\uparrow$1.7} & 22.3
  & 20.5\,{\color{gray}=} & 20.6 \\

\midrule
\multicolumn{17}{l}{\textit{Large models (15B--34B)}} \\[1pt]

StarCoder2-15B
  & 65.9 & 97.6
  & 51.2\,{\color{red}$\downarrow$14.7} & 97.0
  & 54.3\,{\color{red}$\downarrow$11.6} & 95.1
  & 62.8\,{\color{red}$\downarrow$3.1}  & 93.9
  &  6.8 &  6.9
  &  8.0\,{\color{green!60!black}$\uparrow$1.2} &  8.1
  &  7.3\,{\color{green!60!black}$\uparrow$0.5} &  7.4
  &  7.6\,{\color{green!60!black}$\uparrow$0.8} &  7.7 \\

Codestral-22B
  & 72.6 & 97.0
  & 58.5\,{\color{red}$\downarrow$14.1} & 97.6
  & 65.2\,{\color{red}$\downarrow$7.4}  & 96.3
  & 70.7\,{\color{red}$\downarrow$1.9}  & 96.3
  & 23.0 & 23.1
  & 22.5\,{\color{red}$\downarrow$0.5} & 22.6
  & 22.7\,{\color{red}$\downarrow$0.3} & 22.8
  & 22.6\,{\color{red}$\downarrow$0.4} & 22.7 \\

CodeLlama-34B
  & 51.2 & 89.6
  & 42.1\,{\color{red}$\downarrow$9.1} & 89.0
  & 45.1\,{\color{red}$\downarrow$6.1} & 89.6
  & 50.0\,{\color{red}$\downarrow$1.2} & 92.7
  & 13.6 & 13.7
  & 11.6\,{\color{red}$\downarrow$2.0} & 11.7
  & 13.1\,{\color{red}$\downarrow$0.5} & 13.2
  & 12.4\,{\color{red}$\downarrow$1.2} & 12.5 \\

DeepSeek-Coder-33B
  & 72.0 & 97.6
  & 61.6\,{\color{red}$\downarrow$10.4} & 97.0
  & 68.3\,{\color{red}$\downarrow$3.7}  & 97.0
  & 73.8\,{\color{green!60!black}$\uparrow$1.8} & 97.6
  & 20.3 & 20.4
  & 19.4\,{\color{red}$\downarrow$0.7} & 19.5
  & 19.9\,{\color{red}$\downarrow$0.4} & 20.0
  & 18.8\,{\color{red}$\downarrow$1.5} & 18.9 \\

Qwen2.5-Coder-32B
  & 85.4 & 98.8
  & 73.2\,{\color{red}$\downarrow$12.2} & 98.2
  & 84.1\,{\color{red}$\downarrow$1.3}  & 97.0
  & 86.6\,{\color{green!60!black}$\uparrow$1.2} & 99.4
  & 32.0 & 32.1
  & 30.6\,{\color{red}$\downarrow$1.4} & 30.7
  & 31.7\,{\color{red}$\downarrow$0.3} & 31.8
  & 31.2\,{\color{red}$\downarrow$0.8} & 31.3 \\

\midrule
\multicolumn{17}{l}{\textit{Reasoning models (API)}} \\[1pt]

GPT-5-mini
  & 96.3 & 97.6
  & 86.6\,{\color{red}$\downarrow$9.7} & 95.7
  & 89.0\,{\color{red}$\downarrow$7.3} & 91.5
  & 93.3\,{\color{red}$\downarrow$3.0} & 96.3
  & 52.5 & 52.6
  & 48.5\,{\color{red}$\downarrow$4.0} & 48.6
  & 52.5\,{\color{gray}=} & 52.6
  & 53.2\,{\color{green!60!black}$\uparrow$0.7} & 53.3 \\

Claude Sonnet 4
  & 95.7 & 98.2
  & 86.0\,{\color{red}$\downarrow$9.7} & 97.6
  & 89.0\,{\color{red}$\downarrow$6.7} & 97.6
  & 95.7\,{\color{gray}=} & 97.6
  & 51.1 & 51.2
  & 49.6\,{\color{red}$\downarrow$1.5} & 49.7
  & 49.9\,{\color{red}$\downarrow$1.2} & 50.0
  & 50.7\,{\color{red}$\downarrow$0.4} & 50.8 \\

\bottomrule
\end{tabular}
}
\vspace{0.6em}
\end{table*}


\begin{table}[t]
\vspace{-0.6em}
\centering
\small
\caption{Pass@1 (\%) on \textsc{HumanEval};
\textit{LV$_{\text{name}}$} preserves the original function name while paraphrasing the docstring; \textit{LV} applies the full paraphrasing. $\Delta_{\text{vocab}}$ = LV$_{\text{name}}$ $-$ Orig (effect of docstring paraphrasing); $\Delta_{\text{name}}$ = LV $-$ LV$_{\text{name}}$ (additional effect of identifier renaming).}
\vspace{-0.8em}
\label{tab:lv_ablation}
\begin{tabular}{lccccc}
\toprule
\textbf{Model} & \textbf{Orig} & \textbf{LV$_{\text{name}}$} & \textbf{LV} & $\Delta_{\text{vocab}}$ & $\Delta_{\text{name}}$ \\
\midrule
\multicolumn{6}{l}{\textit{Small models}} \\
Qwen2.5-Coder-7B    & 82.3 & 78.0 & 75.6 & $-$4.3 & $-$2.4 \\
DeepSeek-Coder-6.7B & 72.6 & 69.5 & 60.4 & $-$3.1 & $-$9.1 \\
CodeLlama-7B        & 37.2 & 34.8 & 29.3 & $-$2.4 & $-$5.5 \\
\midrule
\multicolumn{6}{l}{\textit{Large models}} \\
Qwen2.5-Coder-32B   & 85.4 & 84.1 & 84.1 & $-$1.3 & $\phantom{-}$0.0 \\
DeepSeek-Coder-33B  & 72.0 & 73.2 & 68.3 & $-$3.7   & $+$1.2    \\
Codestral-22B       & 72.6 & 68.9 & 65.2 & $-$3.7 & $-$3.7 \\
StarCoder2-15B      & 65.9 & 54.9 & 54.3 & $-$11.0 & $-$0.6 \\
CodeLlama-34B       & 51.2 & 41.5 & 45.1 & $-$9.7 & $+$3.6 \\
\midrule
\multicolumn{6}{l}{\textit{API models}} \\
GPT-5-mini          & 96.3 & 90.9 & 89.0 & $-$5.4 & $-$1.9 \\
Claude Sonnet 4     & 95.7 & 96.3 & 89.0 & $+$0.6 & $-$7.3 \\
\bottomrule
\end{tabular}%
\vspace{-0.5em}
\end{table}

\begin{figure*}[t]
    \centering
    \vspace{-0.4em}
    \includegraphics[width=0.8\linewidth]{figures/rq2_attention_orig.png}
    \vspace{-1.0em}
    \caption{Normalized last-layer last-token attention distributed across prompt regions for HumanEval (Signature, Description, Examples) and LiveCodeBench (Description, I/O Format, Sample I/O), averaged over all attention heads. 
    }
    \vspace{-0.7em}
    \label{fig:attention}
\end{figure*}

\subsection{Non-uniform mutation effects}

\begin{table}[t]
\centering
\small
\caption{Per-example flip analysis under LV and US mutations on LCB and HumanEval.
Fail$\to$Pass counts examples failing on the original prompt but passing on the mutated
prompt; Pass$\to$Fail counts the reverse.}
\vspace{-0.8em}
\begin{tabular}{lrrrrrrr}
\toprule
 & \multicolumn{3}{c}{\textbf{LCB}} & \multicolumn{3}{c}{\textbf{HumanEval}} \\
\cmidrule(lr){2-4} \cmidrule(lr){5-7}
\textbf{Model} & \textbf{F$\to$P} & \textbf{P$\to$F} & \textbf{Ratio}
               & \textbf{F$\to$P} & \textbf{P$\to$F} & \textbf{Ratio} \\
\midrule
\multicolumn{7}{l}{\textit{Lexical Vagueness (LV) mutations}} \\
\midrule
Qwen2.5-Coder-32B  & 45 & 48 & 0.94 &  7 &  9 & 0.78 \\
CodeLlama-34B      & 36 & 41 & 0.88 & 16 & 26 & 0.62 \\
DeepSeek-33B       & 36 & 40 & 0.90 & 15 & 21 & 0.71 \\
StarCoder2-15B     & 33 & 28 & 1.18 &  7 & 26 & 0.27 \\
Codestral-22B      & 45 & 48 & 0.94 &  9 & 21 & 0.43 \\
Qwen2.5-Coder-7B   & 52 & 35 & 1.49 &  5 & 16 & 0.31 \\
CodeLlama-7B       & 18 & 17 & 1.06 & 11 & 24 & 0.46 \\
DeepSeek-6.7B      & 32 & 47 & 0.68 &  7 & 27 & 0.26 \\
GPT-4o-mini        & 62 & 62 & 1.00 &  1 & 13 & 0.08 \\
Claude-Sonnet-4    & 52 & 64 & 0.81 &  1 & 12 & 0.08 \\
\midrule
\textit{Average}   & \textit{41.1} & \textit{43.0} & \textit{0.99}
                   & \textit{7.9}  & \textit{19.5} & \textit{0.40} \\
\midrule
\multicolumn{7}{l}{\textit{Under-Specification (US) mutations}} \\
\midrule
Qwen2.5-Coder-32B  & 17 & 31  & 0.55 &  4 & 24 & 0.17 \\
CodeLlama-34B      & 10 & 30  & 0.33 &  6 & 21 & 0.29 \\
DeepSeek-33B       & 20 & 29  & 0.69 &  8 & 25 & 0.32 \\
StarCoder2-15B     & 24 & 12  & 2.00 &  3 & 27 & 0.11 \\
Codestral-22B      & 28 & 33  & 0.85 &  4 & 27 & 0.15 \\
Qwen2.5-Coder-7B   & 27 & 29  & 0.93 &  4 & 28 & 0.14 \\
CodeLlama-7B       & 17 & 14  & 1.21 &  7 & 20 & 0.35 \\
DeepSeek-6.7B      & 32 & 35  & 0.91 &  7 & 32 & 0.22 \\
GPT-4o-mini        & 65 & 105 & 0.62 &  1 & 17 & 0.06 \\
Claude-Sonnet-4    & 54 & 69  & 0.78 &  3 & 19 & 0.16 \\
\midrule
\textit{Average}   & \textit{29.4} & \textit{38.7} & \textit{0.89}
                   & \textit{4.7}  & \textit{24.0} & \textit{0.20} \\
\bottomrule
\end{tabular}
\label{tab:flip_lcb_humaneval}
\vspace{-0.7em}
\end{table}

\begin{table}[t]
\centering
\caption{Cross-model consistent Fail$\to$Pass flip statistics on LCB under LV (lexical vagueness) and US (underspecification). A flip is ``consistent'' if it occurs in at least 2 models. ``Clean'' tasks are those where no model flipped P$\to$F.}
\vspace{-0.7em}
\label{tab:lv_us_f2p_stats}
\begin{tabular}{lrr}
\toprule
\textbf{Statistic} & \textbf{LV} & \textbf{US} \\
\midrule
\multicolumn{3}{l}{\textit{Task-level counts}} \\
\midrule
Tasks with F$\to$P in $\geq$2 models       & 69 & 27 \\
\quad Clean (P$\to$F\,=\,0)               & 33 & 14 \\
\quad Mixed (P$\to$F\,$\geq$\,1)          & 36 & 13 \\
Tasks with F$\to$P in $\geq$3 models       & 15 &  3 \\
Tasks with F$\to$P in $\geq$4 models       &  1 &  0 \\
\midrule
\multicolumn{3}{l}{\textit{Model--task pair counts (consistent tasks only)}} \\
\midrule
Total F$\to$P pairs                        & 154 & 57 \\
Total P$\to$F pairs                        &  50 & 16 \\
Net                                        & $+104$ & $+41$ \\
F$\to$P / P$\to$F ratio                   & 3.08 & 3.56 \\
\midrule
\multicolumn{3}{l}{\textit{Dominant change/constraint type (tasks, \% of section)}} \\
\midrule
Terminology / variable rename              & 33\ \ (48\%) & --- \\
Task description rephrasing               & 27\ \ (39\%) & --- \\
I/O description change                    &  6\ \ \ (9\%) & --- \\
Domain noun substitution                  &  2\ \ \ (3\%) & --- \\
Value range removed                       & --- & 12\ \ (44\%) \\
Size bound removed                        & --- &  8\ \ (30\%) \\
Input format constraint removed           & --- &  4\ \ (15\%) \\
Uniqueness / ordering / other             & --- &  3\ \ (11\%) \\
\bottomrule
\end{tabular}
\vspace{0.5em}
\end{table}

\begin{table}[t]
\centering
\vspace{-0.5em}
\caption{%
  Test-suite coverage of passing solutions for F$\to$P and P$\to$F flip cases on LiveCodeBench (10 models, US/LV mutations).
  \emph{Line coverage}: fraction of executable statements reached by at least one test. 
  \emph{Branch coverage}: fraction of conditional branch outcomes (true/false arms) exercised. 
}
\vspace{-0.7em}
\label{tab:coverage_f2p_p2f}
\resizebox{\linewidth}{!}{%
\setlength{\tabcolsep}{5pt}
\begin{tabular}{l rr rr rr}
\toprule
 & \multicolumn{2}{c}{\textbf{LV}}
 & \multicolumn{2}{c}{\textbf{US}}
 & \multicolumn{2}{c}{\textbf{All}} \\
\cmidrule(lr){2-3}\cmidrule(lr){4-5}\cmidrule(lr){6-7}
\textbf{Metric}
 & \textbf{F$\to$P} & \textbf{P$\to$F}
 & \textbf{F$\to$P} & \textbf{P$\to$F}
 & \textbf{F$\to$P} & \textbf{P$\to$F} \\
\midrule
$N$                             & 411  & 429  & 294  & 387  & 1001 & 1154 \\
Mean line cov.\ (\%)            & 91.8 & 92.7 & 90.8 & 90.8 & 91.2 & 91.9 \\
Median line cov.\ (\%)          & 95.9 & 95.8 & 95.7 & 94.4 & 95.7 & 95.5 \\
Mean branch cov.\ (\%)          & 83.7 & 84.9 & 83.0 & 83.6 & 83.4 & 84.5 \\
\bottomrule
\end{tabular}}
\vspace{-0.5em}
\end{table}

Given that LiveCodeBench shows near-zero net sensitivity in RQ1, is this because mutations have no effect, or because something more nuanced is occurring? Table~\ref{tab:flip_lcb_humaneval} break down aggregate Pass@1 deltas into per-example transitions: problems  that fail on the original prompt but pass on the mutated one (F$\to$P), and the reverse (P$\to$F), revealing that the near-zero net effect on LCB results 
from two opposing trends of comparable magnitude that cancel each other.

\textbf{On HumanEval, mutations produce strongly asymmetric flip distributions dominated by degradation.} As shown in Table~\ref{tab:flip_lcb_humaneval}, under LV, the average F$\to$P/P$\to$F ratio is $0.40$, meaning that for every example a mutation leads to fixes, it breaks 2.5 others.  Under US~\ref{tab:flip_lcb_humaneval}, this ratio drops further to $0.20$, one improvement for every five regressions. Across all models, P$\to$F counts consistently and substantially  exceed F$\to$P counts, confirming that mutations act almost exclusively as a source of degradation on single-specification prompts.

\textbf{On LCB, the same mutations produce balanced flip distributions, with improvements offsetting degradations.} Under LV~\ref{tab:flip_lcb_humaneval}, the average F$\to$P/P$\to$F ratio is $0.99$, near perfect balance, with four models recording a positive net (StarCoder2-15B: $+5$, Qwen2.5-Coder-7B: $+17$, CodeLlama 7B: $+1$). Under US~\ref{tab:flip_lcb_humaneval}, the ratio is $0.89$, still substantially higher than on HumanEval. The near-zero aggregate effect on LCB is therefore not the absence of signal but the result of two competing forces canceling each other out.

\textbf{A subset of LCB problems improves consistently and exclusively across models.} Table~\ref{tab:lv_us_f2p_stats} identifies tasks where F$\to$P flips occur in at least two models with no accompanying P$\to$F flips elsewhere, denoted \textit{clean} tasks. Under LV, 33 such tasks are identified; under US, 14. On these tasks, the F$\to$P/P$\to$F ratio reaches $3.08$ and $3.56$  respectively, with net gains of $+104$ and $+41$ model-task pairs.  The consistency of these improvements across models, rather than being isolated to a single model, suggests that the original prompts contain systematic properties that hinder correct code generation, independent of model-specific behavior. 

To rule out the possibility that these passing solutions are artifacts of insufficient test coverage, Table~\ref{tab:coverage_f2p_p2f} reports branch and line coverage of F$\to$P solutions: mean line coverage reaches 91.2\% and mean branch coverage 83.4\%, both comparable to P$\to$F solutions (91.9\% and 84.5\% respectively), confirming that the test suite exercises the generated code thoroughly and that the observed improvements reflect correctness gains. \\

\begin{tcolorbox}[colback=gray!10, colframe=gray!50, boxrule=0.5pt]
\textbf{ Takeaway.} 
Under specified prompts do not uniformly degrade code generation. On LiveCodeBench, 47 problems improve consistently and exclusively across multiple models when vagueness or under-specification is introduced, revealing that for a non-trivial subset of real-world problems, over-specification silently misleads models in ways not apparent from the
prompt itself.
\end{tcolorbox}

\subsection{Causes of improvement}

Having established that a consistent subset of LCB problems improves under vagueness and under-specification, we now ask what linguistic mechanisms drive these improvements and what they reveal about how models process specifications.

\begin{figure*}[t]
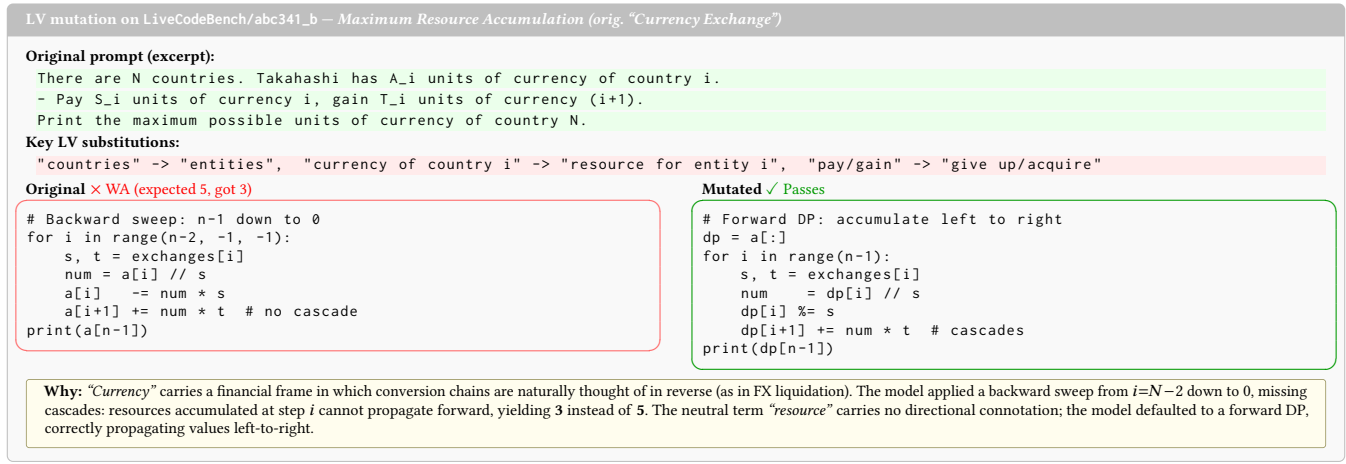

\centering
\scriptsize
\vspace{-0.5em}
\begin{tcolorbox}[
    colback=gray!5, colframe=gray!50,
    title={LV mutation on \texttt{LiveCodeBench/abc341\_b}
           --- \textit{Maximum Resource Accumulation (orig.\ ``Currency Exchange'')}},
    fonttitle=\scriptsize\bfseries,
    boxrule=0.4pt, arc=2pt,
    left=4pt, right=4pt, top=2pt, bottom=2pt
]
\textbf{Original prompt (excerpt):}
\begin{lstlisting}[basicstyle=\scriptsize\ttfamily, backgroundcolor=\color{green!8},
                   frame=none, xleftmargin=4pt, aboveskip=1pt, belowskip=1pt]
There are N countries. Takahashi has A_i units of currency of country i.
- Pay S_i units of currency i, gain T_i units of currency (i+1).
Print the maximum possible units of currency of country N.
\end{lstlisting}
\textbf{Key LV substitutions:}
\begin{lstlisting}[basicstyle=\scriptsize\ttfamily, backgroundcolor=\color{red!8},
                   frame=none, xleftmargin=4pt, aboveskip=1pt, belowskip=2pt]
"countries" -> "entities",  "currency of country i" -> "resource for entity i",  "pay/gain" -> "give up/acquire"
\end{lstlisting}
\begin{minipage}[t]{0.48\linewidth}
\textbf{Original} \textcolor{red}{$\times$ WA (expected 5, got 3)}
\begin{lstlisting}[basicstyle=\scriptsize\ttfamily, frame=single,
                   frameround=tttt, rulecolor=\color{red!60},
                   aboveskip=1pt, belowskip=0pt]
# Backward sweep: n-1 down to 0
for i in range(n-2, -1, -1):
    s, t = exchanges[i]
    num = a[i] // s
    a[i]   -= num * s
    a[i+1] += num * t  # no cascade
print(a[n-1])
\end{lstlisting}
\end{minipage}
\hfill
\begin{minipage}[t]{0.48\linewidth}
\textbf{Mutated} \textcolor{green!60!black}{$\checkmark$ Passes}
\begin{lstlisting}[basicstyle=\scriptsize\ttfamily, frame=single,
                   frameround=tttt, rulecolor=\color{green!60!black},
                   aboveskip=1pt, belowskip=0pt]
# Forward DP: accumulate left to right
dp = a[:]
for i in range(n-1):
    s, t = exchanges[i]
    num    = dp[i] // s
    dp[i] %= s
    dp[i+1] += num * t  # cascades
print(dp[n-1])
\end{lstlisting}
\end{minipage}
\begin{tcolorbox}[colback=yellow!8, colframe=yellow!50!black,
                  boxrule=0.3pt, arc=1pt, left=4pt, right=4pt,
                  top=1pt, bottom=1pt, before upper={\vspace{-2pt}}]
\scriptsize
\textbf{Why:} \textit{``Currency''} carries a financial frame in which conversion chains are naturally
thought of in reverse (as in FX liquidation). The model applied a backward sweep from $i{=}N{-}2$
down to $0$, missing cascades: resources accumulated at step $i$ cannot propagate forward, yielding
\textbf{3} instead of \textbf{5}. The neutral term \textit{``resource''} carries no directional
connotation; the model defaulted to a forward DP, correctly propagating values left-to-right.
\end{tcolorbox}
\end{tcolorbox}
\vspace{-1.0em}
\caption{Example of an LV mutation improving performance by neutralizing a domain-specific vocabulary cue.}
\label{fig:lv_motivating_example}
\vspace{-0.5em}
\end{figure*}

\textbf{Under lexical vagueness, improvements are driven primarily by disrupting over-fitted retrieval cues.} Table~\ref{tab:root-causes} shows that the two dominant mechanisms in LCB under LV are variable renaming that breaks overfitted retrieval (27\%) and terminology generalization that activates better algorithmic patterns (27\%). In both cases, the original prompt contains domain-specific vocabulary or identifier names that prime the model toward a memorized but incorrect solution; replacing them with vaguer alternatives forces the model to reason from the problem structure instead (Figure~\ref{fig:lv_motivating_example}). A further 13\% of cases involve function renaming, in which rewriting under a new signature incidentally corrects parsing or boundary bugs as a side effect. The remaining cases involve softening algorithmic language, which frees the model to select simpler correct strategies over over-fitted templates (10\%), rephrasing input descriptions, which triggers regeneration of parsing code and corrects I/O handling bugs (10\%), and constraint rewording that disambiguates boundary conditions left unclear in the original phrasing (8\%).

\textit{Under under-specification, constraint removal disrupts two distinct failure modes.}
Table~\ref{tab:us_root_causes} identifies constraint-triggered wrong input parsing as the dominant mechanism (39\%): specific constraint lines prime the model to write LeetCode-style I/O handling incompatible with the actual stdin format, and removing them forces a correct rewrite. The second mechanism, constraint-anchored wrong algorithm (35\%), is equally systematic: numerical bounds activate memorized algorithm shortcuts that are misapplied to the problem at hand, and removing the bound allows the model to select a general correct strategy (Figure~\ref{fig:us_motivating_example}). Together, these two mechanisms account for 74\% of US improvement cases. In a further 17\% of cases, removing a bound simplifies the problem sufficiently that the model defaults to a fully general implementation, forward DP, full BFS, direct enumeration, that is structurally harder to get wrong than the optimized solution the original constraint had scoped it toward.

These findings carry direct implications: for practitioners, reducing specification precision can sometimes improve generation; for benchmark designers, completeness of specification does not guarantee prompt quality; and for model developers, sensitivity to retrieval cues points to a generalisation gap that scale alone does not resolve.

\begin{mdframed}[style=takeawaystyle]
\textbf{Takeaway.} 
Code generation improvements occur when original prompts contain retrieval cues, domain terms, identifier names, or constraint lines that trigger memorized but incorrect solution strategies. Removing or softening these cues forces models to reason from the problem structure, often arriving at correct solutions via simpler algorithms. Specification completeness does not guarantee prompt quality, and retrieval-cue sensitivity reveals a generalization gap that model scale alone does not resolve.
\end{mdframed}

\begin{table*}[ht]
\small
\centering
\renewcommand{\arraystretch}{1.3}
\caption{Root-cause analysis of LV prompt mutations inducing behavioral change in Qwen-7B
         (incorrect $\to$ correct) on LiveCodeBench ($n=52$).}
\label{tab:root-causes}
\begin{tabular}{p{3.0cm}p{0.75cm}p{0.6cm}
                p{\dimexpr\linewidth-3.0cm-0.75cm-0.6cm-8\tabcolsep\relax}}
\toprule
\textbf{Root Cause} & \textbf{n} & \textbf{\%} & \textbf{Interpretation} \\
\midrule
\rowcolor[RGB]{220,235,255}
Variable rename breaks overfit &
14 & 27 &
Renaming key variables (e.g.\ \texttt{s}$\to$\texttt{seq}, \texttt{n,x}$\to$\texttt{a,b}) prevents retrieval of a memorised incorrect solution, forcing reasoning from the problem description. \\
\rowcolor[RGB]{220,255,220}
Terminology generalisation triggers better patterns &
14 & 27 &
Replacing domain nouns with generic terms (``graph''$\to$``network'', ``boxes''$\to$``containers'') activates canonical algorithmic templates (BFS, knapsack, set-union) over problem-specific shortcuts. \\
\rowcolor[RGB]{240,220,255}
Function rename changes approach &
7 & 13 &
An explicit function signature (e.g.\ \texttt{gather\_pairs(n)}) anchors code structure, incidentally fixing ancillary bugs (wrong parsing, boundary errors) by forcing a full rewrite. \\
\rowcolor[RGB]{255,245,200}
Vague algorithm frees model &
5 & 10 &
Softening algorithmic language (``find minimum''$\to$``smallest possible'') lets the model choose a simpler correct strategy (brute-force, enumerate-all) over an overfitted template (broken DP, wrong greedy). \\
\rowcolor[RGB]{255,225,210}
Input-format change fixes parsing &
5 & 10 &
Rephrasing the input description triggers regeneration of parsing code, correcting token-iteration bugs, index offsets, or \texttt{sys.stdin.read()} misuse. \\
\rowcolor[RGB]{255,235,200}
Constraint rephrasing clarifies &
4 & 8 &
Explicit constraint wording (``divisors smaller than $x$'', ``no empty spots'') disambiguates boundaries left unclear in the original, preventing off-by-one errors. \\
\rowcolor[RGB]{230,230,230}
Other (stochastic/unclear) &
3 & 6 &
Surface-form change shifts output stochastically onto a correct solution; no single linguistic feature is identifiable as responsible. \\
\bottomrule
\end{tabular}
\vspace{-0.5em}
\end{table*}

\begin{table*}[ht]
\small
\centering
\renewcommand{\arraystretch}{1.3}
\caption{Taxonomy of underspecification (US) mutations that induced a Fail$\to$Pass flip on Claude-Sonnet-4 / LiveCodeBench (54 instances). Rows describe the mechanism by which removing a constraint steered the model toward a correct solution.}
\vspace{-0.5em}
\label{tab:us_root_causes}
\begin{tabular}{p{3.2cm}p{0.6cm}p{0.6cm}p{3.2cm}
                p{\dimexpr\linewidth-3.2cm-0.6cm-0.6cm-3.2cm-10\tabcolsep\relax}}
\toprule
\textbf{Root Cause} & \textbf{N} & \textbf{\%} & \textbf{Constraint Types} & \textbf{Mechanism \& Example} \\
\midrule
\rowcolor[RGB]{220,235,255}
I/O format primed by constraint text &
21 & 39 &
value\_range (7), size\_bound (5), input\_format (4), uniqueness (3), other (2) &
Seeing a constraint line (e.g.\ \texttt{1\,$\leq$\,nums[i]\,$\leq$\,10\^{}5}) primed the model to use LeetCode-style parsing (\texttt{input().split()}) rather than bracket-formatted competitive-programming stdin. Removing the line eliminated the priming cue; the algorithm itself was never wrong. \\
\rowcolor[RGB]{220,255,220}
Constraint anchored a wrong algorithm &
19 & 35 &
value\_range (11), size\_bound (5), input\_format (2), other (1) &
A numeric or categorical bound triggered a memorised shortcut that did not fit the problem: e.g.\ \texttt{C\textsubscript{i}\,$\leq$\,10\^{}9} activated a star-graph diameter formula instead of a two-BFS approach; \textit{M is prime} invoked modular-inverse arithmetic instead of a general DP. Without the cue, the model fell back to the correct general algorithm. \\
\rowcolor[RGB]{255,245,200}
Bound pushed model into a buggy optimisation &
9 & 17 &
value\_range (3), size\_bound (2), input\_format (2), other (2) &
The constraint scoped the model toward an optimised solution with a structural bug (wrong DP dimension order, premature greedy pruning). Freed from the bound, the model chose a simpler, fully general implementation (plain forward DP, full enumeration) that was harder to get wrong. \\
\rowcolor[RGB]{240,230,240}
No identifiable causal link &
4 & 7 &
size\_bound (2), input\_format (1), other (1) &
The deleted text was incidental (a minor size cap or formatting note); no direct linguistic trigger is apparent. The improvement is best attributed to re-sampling variance. \\
\rowcolor[RGB]{255,225,210}
Constraint triggered overfit template &
1 & 2 &
value\_range (1) &
Domain vocabulary in the constraint surface-matched a memorised but incorrect solution template. Removing it allowed the model to reason from problem structure rather than retrieve a mismatched pattern. \\
\bottomrule
\end{tabular}
\end{table*}


%% file: Related_work.tex
\section{Related work}
\subsection{LLM-based code generation and evaluation}

Code generation benchmarks have evolved from static, single-language test suites to more rigorous and diverse evaluation frameworks. Earlier benchmarks such as \textit{HumanEval}~\cite{humaneval} and \textit{MBPP}~\cite{mbpp} established the standard paradigm for evaluating functional correctness using unit-test pass rates (e.g., \textit{pass@k}). However, subsequent work has shown that these evaluations can overestimate the model's performance. For example, \textit{EvalPlus}~\cite{liu2023evalplus} increases HumanEval with substantially more test cases, demonstrating that increasing test coverage can significantly reduce reported scores and expose weaknesses in generated solutions.

More recent benchmarks have sought to address additional limitations of earlier evaluations. \textit{LiveCodeBench}~\cite{livecodebench2024} mitigates benchmark contamination by continuously introducing new competitive-programming problems, while \textit{BigCodeBench}~\cite{zhuo2025bigcodebench} extends evaluation to more complex, library-intensive tasks that require diverse API usage and more realistic programming workflows.

Alongside these benchmarks, the ecosystem of code-oriented language models has expanded rapidly \cite{BrownMRSKDNSSAA20}. Open-weight models include \textit{Code Llama}~\cite{roziere2023codellama}, \textit{StarCoder2}~\cite{lozhkov2024starcoder2}, \textit{DeepSeek-Coder}~\cite{guo2024deepseek}, and \textit{Qwen2.5-Coder}~\cite{hui2024qwen25coder}, while proprietary systems such as GPT-5~\cite{openai2025gpt5} and Claude Code~\cite{anthropic2024claude} represent the state of the art in commercial deployments. Despite the growing diversity of benchmarks and models, evaluation practices remain largely focused on correctness under clean, well-specified prompts,  leaving model behavior under realistic prompt imperfections unexplored.

\subsection{Robustness of code LLMs to prompt permutations}

The most closely related line of work studies how prompt perturbations affect code generation quality. \textit{ReCode}~\cite{recode} introduced over 30 semantic-preserving transformations, variable renaming, dead-code insertion, and docstring reformatting on HumanEval~\cite{humaneval} and MBPP~\cite{mbpp}, revealing substantial performance degradation. Mastropaolo et al.~\cite{mastropaolo2023robustness} show that semantically equivalent Javadoc paraphrases alter Copilot output in nearly half of cases. \textit{NLPerturbator}~\cite{NLPerturbator} extended  this to 18 natural-language variation categories derived from real developer data, reporting drops of up to 21.2\%, while Rabbi et al.~\cite{rabbi2025multilanguage} broadened the analysis to multiple programming languages. Sclar et al.~\cite{formatspread} demonstrated that formatting changes alone cause accuracy swings of up to 76\%. More recently, Larbi et al.~\cite{larbi2025prompts} examined ambiguous, incomplete, and contradictory 
task descriptions in a restricted set of models and reported large performance drops for all models and defects. 

While these studies collectively establish that code LLMs are sensitive to prompt perturbations, they share a common limitation: all evaluate robustness exclusively on minimal, single-specification benchmarks and frame perturbations as purely degrading. 

\subsection{Prompt under-specification}

A parallel line of work addresses the consequences of ambiguous or incomplete specifications in code generation. ClarifyGPT~\cite{mu2024clarifygpt} detects ambiguous requirements and improves Pass@1 through targeted clarification, while TiCoder~\cite{TiCoder} proposed test-driven interactive  clarification to reduce intent mismatch. SpecFix~\cite{SpecFix} automated ambiguity repair, finding that 43.58\% of benchmark descriptions contain actionable ambiguity. These works establish that specification imperfection is pervasive and consequential, but they focus exclusively on repairing ambiguity, treating all specification degradation as harmful.  The assumption that more specification is always better has been questioned in adjacent work. Yang et al.~\cite{yang2025whatprompts} found that over-specified prompts can overwhelm general-purpose LLMs, identifying over-specification as an anti-pattern. Döderlein et al.~\cite{doderlein2025piloting} observed that removing prompts entirely can sometimes improve code generation, and Break-the Chain~\cite{charalambous2025breakthechain} showed that narrative reframing of  coding problems improves accuracy by up to 35.3\% for some models. 

However, these observations are scattered: none provides a controlled comparison across benchmark types, none quantifies improvement consistency across models, and none identifies the linguistic mechanisms responsible. Our work provides the first systematic account of when and why reducing specification precision improves code generation, grounded in root-cause analysis across 47 consistently improved problems.

%% file: threats.tex
\section{Threats to Validity}
Our mutations are produced automatically by \texttt{gpt-5-mini}, which may shape them in ways that reflect the generator's own tendencies~\cite{llmjudge,WangLCCZLL24}. Although the judge and manual checks confirm high quality, some variants may still be less natural than what real developers would introduce~\cite{JustJIEHF14}.
HumanEval is known to overlap with model training data~\cite{yang2023rethinking,MattonSATA24}, so some models may recall solutions rather than reason from the prompt, making our measured drops a lower bound. Both benchmarks are Python-only, and results may not hold for other languages~\cite{rabbi2025multilanguage} or for tasks that require heavy library use~\cite{BigCodeBench}, where the specification structure plays a different role.
We use Pass@1 with greedy decoding to keep results reproducible, but this hides the variance that comes with sampling~\cite{SongWLXL24,OuyangZHW24} and underestimates what models can do with multiple attempts~\cite{Chen2021}. Small individual flips may therefore reflect output variance rather than true sensitivity to the prompt change.
Finally, our exploratory analysis ~\cite{GonzalezPDL23,RalphABBDDE21} forms a first attempt to identify what impacts the LLM-based code generation correctness. Therefore, the observed improvements could, in principle, be coincidental or reflect confounding factors. We mitigate this by requiring consistency across multiple models, verifying test-suite coverage of flipped solutions, and relying on human annotation, but causal confirmation requires future controlled experiments. To our knowledge, this is the first study to examine these mechanisms at this level of granularity; as an exploratory study, it is, by definition, hypothesis-generating rather than hypothesis-confirming, and our findings should be interpreted accordingly.

%% file: conclusion.tex
\section{Conclusion}

This study investigated how prompt structure, task complexity, and specification richness relates to LLM robustness to prompt mutations. Our results show that robustness is not an intrinsic model property but an artifact of prompt structure: the same mutant types that cause substantial Pass@1 drops on HumanEval (US -11.8\%, LV -7.1\%) produce near-zero net effects on LiveCodeBench, where multi-layered specifications provide redundant specification layers. Crucially, this near-zero effect masks two opposing forces of comparable magnitude, improvements and degradations, which cancel rather than being absent. On 47 consistently improved LCB problems, root-cause analysis identified retrieval-cue disruption as the dominant mechanism: domain-specific vocabulary, identifier names, and constraint lines prime models toward memorized but incorrect solutions, and their removal forces reasoning from problem structure. 

These findings suggest that practitioners should be careful with over-specified prompts that may silently mislead models. Benchmark designers may consider that minimal-specification benchmarks probably do not generalize to richer task formats, and model developers should consider that words triggering retrieval cues may well hamper the effectiveness of the model reasoning.
